\documentclass[a4paper,11pt,oneside]{article}

\usepackage[english]{babel}
\usepackage[utf8]{inputenc}
\usepackage[T1]{fontenc}
\usepackage{amsmath,amssymb,amsfonts,amsthm}
\usepackage{fixltx2e} % fixes some nasty thins about latex
\usepackage[svgnames]{xcolor}
\usepackage{hyperref} 
\usepackage[hyperpageref]{backref}
\hypersetup{bookmarksnumbered,linktocpage,colorlinks,debug, % debug for extra diagnostic messages in log file
pdftitle={On Lorentz-Violating Quatum Field Theories},
pdfauthor={Diego Redigolo},
pdfsubject={Lorentz-Violating Quantum Field Theories},
pdflang={Anglais},
citecolor=[rgb]{0,.7,.1},
urlcolor=blue} 
\RequirePackage[hyperpageref]{backref}  
\usepackage{marginnote}
\usepackage[totalwidth=460truept,totalheight=600truept]{geometry}
\usepackage{latexsym,amssymb,amsmath,graphicx,accents,eucal,slashed,subfigure,axodraw4j}

\begin{document}

\thispagestyle{empty}
\setcounter{page}{1}

\vspace{10mm}

\begin{center} {\textbf{\Large On Lorentz-Violating Supersymmetric Quantum Field Theories}}

\vspace{1.5cm}

Diego \textsc{Redigolo}$^{\,1}$ 

\footnotesize
\vspace*{40pt}

${}^1${\em Universit\'e Libre de Bruxelles (ULB) \& International Solvay Institutes (ISI),\\
Campus Plaine CP231, B-1050, Belgium}

\vspace*{35pt}

\href{mailto:dredigol@ulb.ac.be}{\color{Indigo}\texttt{dredigol@ulb.ac.be}}\texttt{\color{Indigo}}\,
\end{center}
\vspace*{40pt}

\centerline{\textbf{Abstract}}
\vspace*{15pt}

\noindent
We study the possibility of constructing Lorentz-violating supersymmetric quantum field theories under the assumption that these theories have to be described  by lagrangians which are renormalizable by weighted power counting. Our investigation starts from the observation that at high energies Lorentz-violation and the usual supersymmetry algebra are algebraically compatible. Demanding linearity of the supercharges we see that the requirement of renormalizability drastically restricts the set of possible Lorentz-violating supersymmetric theories. In particular, in the case of supersymmetric gauge theories the weighted power counting has to coincide with the usual one and the only Lorentz-violating operators are introduced by some weighted constant $c$ that explicitly appears in the supersymmetry algebra. This parameter does not renormalize and has to be very close to the speed of light at low energies in order to satisfy the strict experimental bounds on Lorentz violation. The only possible models with non trivial Lorentz-violating operators involve neutral chiral superfields and do not have a gauge invariant extension. We conclude that, under the assumption that high-energy physics can be described by a renormalizable Lorentz-violating extension of the Standard Model, the Lorentz fine tuning problem does not seem solvable by the requirement of supersymmetry.
\newpage 
\pagestyle{plain}

\tableofcontents

\section{Introduction}
Lorentz symmetry is a basic ingredient of the Standard Model and it has been verified with great precision by different experiments \cite{Kostelecky:2008ts}. However, from a theoretical point of view it would be possible to violate Lorentz invariance at very high energies requiring this symmetry to be restored at low energies \cite{Chadha:1982qq}. Moreover, it has been shown that Lorentz violation could emerge as a preferred-frame effect both in string theory \cite{Kostelecky:1988zi, Mavromatos:2007xe} and in loop quantum gravity \cite{Gambini:1998it} .\newline
\indent In our approach we want to preserve the requirement of renormalizability as an a priori selection criterion for physical theories at high energies and at the same time enlarge the class of renormalizable theories relaxing the hypothesis of Lorentz invariance but preserving both locality and unitarity.  
\newline\indent The issue of renormalizability in the context of Lorentz-violating theories has been defined and widely studied for both matter and gauge fields \cite{Anselmi:2008ry, Anselmi:2008bq, Anselmi:2008bs, Anselmi:2007ri,Dhar:2009dx, Dhar:2009am}. The basic idea is that one can break Lorentz symmetry at high energies introducing a weighted power counting that weights differently time and space coordinates. Thanks to this modified power counting we can improve the UV behavior of propagators by introducing higher space derivatives terms in the kinetic lagrangian. The number of higher space derivatives in a given theory is parametrized by some integer $n$. This procedure preserves perturbatively the unitarity of the theories because the weighted power counting coincides with the usual one on the time coordinate. 
Lorentz-violating theories become less divergent and  contain new interactions which are non-renormalizable by the usual power counting but renormalizable by the weighted one. We assume that Lorentz-violating terms arise  at energies greater than a breaking energy scale $\Lambda_{L}$.\newline
\indent The implications of Lorentz violation on physics beyond the Standard Model and in particular its consequences on low-energy phenomenology has been widely explored in the literature \cite{Bluhm:2005uj, Colladay:1998fq}. There are several purposes of possible Lorentz-violating extensions of the Standard Model that answer some questions about neutrino physics such as neutrino oscillations and neutrino masses \cite{Anselmi:2008bt} and give an alternative framework which concerns the origin of the mass of elementary particles \cite{Anselmi:2007zz, Anselmi:2009vz, Anselmi:2011ac}.\newline
\indent The recovery of Lorentz invariance in the low-energy limit is not automatic and depends mostly on the running of the coupling constants associated to non-covariant operators that are renormalizable in the usual sense and for this reason are not suppressed by any powers of $\Lambda_{L}$. It has been shown that these coupling constants go to zero in the IR limit for CPT-preserving operators in the context of pure Yang-Mills theories, but grow for CPT-violating ones \cite{Colladay:2006rk}. However, even if we assume CPT invariance we need a strong fine tuning on dimension 2 and dimension 3 operators at low energies to be in agreement with the strict constraints on Lorentz violation \cite{Coleman:1998ti, Iengo:2009ix}. This problem is known in the literature as the Lorentz fine tuning problem \cite{Collins:2004bp}. One of the best candidates for trying to solve this new naturalness problem in the context of Lorentz-violating theories  seems to be supersymmetry \cite{Berger:2001rm, Jain:2005as}. \newline  
\indent On this basis it can be interesting to study quantum field theories that are exact representations of the SUSY algebra and contain Lorentz-violating terms. The idea of combining supersymmetry and Lorentz violation arises from a simple observation: even though the SUSY algebra is constructed as the graded extension of the Poincar\'e algebra, Lorentz violation and the SUSY algebra are not incompatible. In fact, if we weight differently space and time we have to eliminate from the Poincar\'e algebra the generators of the boosts, but these generators do not appear in the anticommutator of two supercharges. Hence we can eliminate all the relations that contain them obtaining again a closed algebra \cite{Berger:2001rm}.\newline
\indent From this algebraic compatibility we can start to construct $N=1$ supersymmetric field theories with Lorentz violation. The classification of Lorentz-violating effective field theories that are exact representations of the SUSY algebra and the study of their quantum corrections have already been the subject of careful investigations in the literature \cite{GrootNibbelink:2004za, Bolokhov:2005cj}. However, in this work we want to focus our attention on the possibility of constructing Lorentz-violating supersymmetric theories which are renormalizable by weighted power counting. If we require the existence of superspace and the linearity of supercharges in the momenta we find, using weighted power counting, that the K\"ahler potential must have the same form as in the Lorentz invariant case and in particular it cannot contain higher space derivatives. This happens because the Grassmannian spinor variables associated with the supercharges are not affected by the weighted power counting. However, for theories with even $n$ we can introduce an even number of higher space derivatives as ``mass'' terms in the superpotential, preserving the rotational invariance in the spatial submanifold $\bar{M}_{\bar{d}}$. In this approach we can classify all the possible theories for chiral superfields and derive the corresponding Feynman rules. The structure of Feynman graphs in the Grassman variables is not modified so that the non-renormalization theorem works as in the Lorentz invariant case.\newline\indent 
Using a result of \cite{GrootNibbelink:2004za}, we show that it is not possible to construct a gauged version of the terms with higher space derivatives in the superpotential. 
Moreover, by looking at the gauge sector we find that a generic Lorentz-violating gauge theory, defined once we have assigned weights to the fields, admits a supersymmetric extension if and only if the weighted power counting coincides with the usual one. Therefore, the requirement of supersymmetry rules out the possibility of having high energy renormalizable extensions of the Standard Model with higher dimensional operators. The only Lorentz-violating operators in the theory have to be renormalizable in the usual sense and their coupling constants need to fulfill the strict experimental bounds on Lorentz violation. In the supersymmetric case the Lorentz invariance recovery is regulated by the constant $c$ in the spatial part of the kinetic term that does not renormalize if we assume that supersymmetry is softly broken at a scale $\mu_{s}<<\Lambda_{L}$. Therefore, the experimental bounds at low energies determine the values of the Lorentz-violating parameter also at high energies and the Lorentz violation effects will be irrelevant. 
The only non trivial high energy theories that we can construct involve neutral superfields which are singlets with respect to the Standard Model gauge group and interact through Landau-Ginzburg vertices $\Phi^{N}$. \newline\indent
However, the problem of constructing Lorentz-violating supersymmetric field theories remains an open field of research because in principle we could consider various modified supersymmetry algebras in which the anticommutator of two conjugate supercharges is linear in the time momentum but non linear in the space momenta. These structures are still compatible with the weighted power counting and the Coleman-Mandula theorem \cite{Coleman:1967ad, Haag:1974qh} and in the limit $\Lambda_{L}\to\infty$ they approach the usual supersymmetry algebra. It is straightforward to find free lagrangians that are invariant under the action of the new supercharges but, however, the non-linearity in the spatial momenta makes the problem of constructing interacting theories a very complicated one.\newline 
\indent This paper is organized as follows. In section $2$ we study the possible supersymmetry algebras in the Lorentz-violating case and discuss the problem of constructing interacting theories when the supercharges are non-linear in the momenta. In particular we explain on which point we disagree with the results of \cite{Xue:2010ih}. In section $3$, assuming linearity of supercharges in the momenta, we construct the most general renormalizable Lorentz-violating theory for chiral superfields in four dimensions. We study the renormalization properties of these theories and show that the usual non-renormalization theorem for supersymmetric theories has a trivial extension to the Lorentz-violating case. We also give a complete classification of the possible supersymmetric Lorentz-violating theories for neutral chiral superfields.  In section $4$ we study the problem of constructing Lorentz-violating gauge theories. We apply a result of \cite{GrootNibbelink:2004za} to show that the theories of section $3$ are not generalizable to the case of charged chiral superfields. Finally, we show that if we demand supersymmetry for gauge theories the weighted power counting has to coincide with the usual one. In section $5$ we discuss the low-energy limit of our theories and the recovery of Lorentz invariance. In the superfield formalism it is almost trivial to show that the Lorentz-violating parameter $c$ in the kinetic lagrangian does not renormalize to all orders of perturbation theory. However, we will explicitly see in components how the requirement of supersymmetry remarkably changes the one loop renormalization group equations for Lorentz-violating theories at low energies. Moreover, we discuss whether or not the deviation from the speed of light is physically observable in the Lorentz-violating supersymmetric theories.
Section $6$ contains our conclusions.

\section{SUSY algebra vs Lorentz symmetry breaking}
\subsection{Berger-Kosetelck\'y construction}
We consider the spacetime manifold $M_{d}$ and we take $d=4$. Defining the weighted power counting, we split $M_{d}$ into the product of two submanifolds $M_{\hat{d}}\times M_{\bar{d}}$ and so the complete Lorentz symmetry $SO(1,3)$ breaks into the residual symmetry $SO(1,\hat{d}-1)\times SO(\bar{d})$. The weighted power counting is defined once we assign the scaling laws, or equivalently the weights of  coordinates or momenta \cite{Anselmi:2007ri}:
\begin{equation}
\hat{x}\rightarrow\hat{x}e^{-\Omega}\ ,\ \ \bar{x}\rightarrow e^{-\Omega/n}\ \ \ \ \ \ ,\ \ \ \ \ \ \ \left[\hat{p}\right]=1\ ,\ \ \left[\bar{p}\right]=\frac{1}{n}\ .\label{1}
\end{equation}
Since we are breaking Lorentz symmetry, we have to discard the boost generator $J_{\hat{\mu}\bar{\nu}}$ between  $M_{\hat{d}}$ and $M_{\bar{d}}$ from the Poincar\'{e} algebra $\mathfrak{p}$. 
However, $J_{\hat{\mu}\bar{\nu}}$ appears only on the left of the  SUSY algebra commutation relations, so we can eliminate from the SUSY algebra $\mathfrak{s}\mathfrak{p}$ the relations that contain  $J_{\hat{\mu}\bar{\nu}}$ obtaining again a closed superalgebra $\mathfrak{s}\mathfrak{p}^{\prime}$.
In general the anticommutator between the two conjugate supercharges will be: 
\begin{equation}\label{commutator}
\left\lbrace Q_{\alpha},\bar{Q}_{\dot{\alpha}}\right\rbrace= 2a\sigma^{\hat{\mu}}_{\alpha\dot{\alpha}}P_{\hat{\mu}}+2b\sigma^{\bar{\mu}}_{\alpha\dot{\alpha}}P_{\bar{\mu}}\ .
\end{equation}
Evaluating (\ref{commutator}) on a generic physical state $\left\vert\psi\right\rangle$ and contracting the spinor indices we obtain
\begin{equation*}
0<\left\langle\psi\right\vert \theta^{\alpha}\lbrace Q_{\alpha},Q_{\dot{\alpha}}\rbrace\theta^{\dot\alpha}\left\vert\psi\right\rangle= 2\left\langle\psi\right\vert\theta^{\alpha}\left(a\sigma^{\hat{\mu}}_{\alpha\dot{\alpha}}P_{\hat{\mu}}+b\sigma^{\bar{\mu}}_{\alpha\dot{\alpha}}P_{\bar{\mu}}\right)\theta^{\dot{\alpha}}\left\vert\psi\right\rangle\ . 
\end{equation*}
The operator $\sigma^{\mu}P_{\mu}$ is positive definite on the physical states and this implies that $a>0\ \forall b$ for both massive and massless particles. 
We can rescale the supercharges defining $Q_{\alpha}^{\prime}=\sqrt{a}Q_{\alpha}$ and dropping the primes we obtain
\begin{equation}\label{eq.slv0}
\left\lbrace Q_{\alpha},\bar{Q}_{\dot{\alpha}}\right\rbrace= 2\sigma^{\hat{\mu}}_{\alpha\dot{\alpha}}P_{\hat{\mu}}+2c\sigma^{\bar{\mu}}_{\alpha\dot{\alpha}}P_{\bar{\mu}}\ . 
\end{equation}
To be consistent with the definition of weighted power counting (\ref{1}) the constant $c$ has to be dimensionless and to weight $\left[c\right]=1-1/n$. The weight of supercharges is fixed by the commutation relations and is equal to their dimension:
\begin{equation*}
\left[Q\right]=\left[\bar{Q}\right]=\frac{1}{2}\ . 
\end{equation*}
We can reabsorb the modification of the SUSY algebra by redefining the Minkowski metric $\eta_{\mu\nu}$ and the $\sigma$ matrices \cite{Berger:2001rm}:
\begin{align}
&\sigma^{\prime\hat{\mu}}=\sigma^{\hat{\mu}}\ ,\ \ \ \ \sigma^{\prime\bar{\mu}}=c\sigma^{\prime\bar{\mu}}\ ,\notag\\
&\left(\sigma^{\prime\mu}\bar{\sigma}^{\prime\nu}+\sigma^{\prime\nu}\bar{\sigma}^{\prime\mu}\right)^{\alpha}_{\beta}= -2\delta^{\alpha}_{\beta}\eta^{\prime\mu\nu}=-2\delta^{\alpha}_{\beta}\left(\eta_{\hat{\mu}\hat{\nu}}+c^{2}\eta_{\bar{\mu}\bar{\nu}}\right)\ ,\notag\\
&{\Box}^{\prime}=\eta^{\prime\mu\nu}\partial_{\mu}\partial_{\nu}=\partial^{\hat{\mu}}\partial_{\hat{\mu}}+c^{2}\partial^{\bar{\mu}}\partial_{\bar{\mu}}\ ,\notag\\
& p^{\prime2}=\eta^{\prime\mu\nu}p_{\mu}p_{\nu}=p^{\hat\mu}p_{\hat\mu}+c^{2}p^{\bar\mu}p_{\bar\mu}\label{pprimo} 
\end{align}
and after these redefinitions the algebra (\ref{eq.slv0}) looks like the usual one. Therefore the structure of the superspace is identical to the Lorentz invariant case as was observed in \cite{Berger:2003ay}. In particular we can define, as usual, the covariant derivatives so that $\left\lbrace D,Q\right\rbrace=\left\lbrace D,\bar{Q}\right\rbrace=0$.
In our construction the crucial requirement is the linearity of the supercharges in the space momenta $P_{\bar{\mu}}$. This assumption makes possible the usual definitions of the  $N=1$ superspace as a coset space defined by the set of variables $z^{\pi}=(\hat{x},\bar{x},\theta,\bar{\theta})$. The operator $U(\hat{y},\bar{y}, \eta, \bar{\eta})=\text{e}^{i(\hat{y}\hat{P}+\bar{y}\bar{P}+i\eta Q+i\bar{\eta}\bar{Q})}$ is a well defined translation in superspace and we can define a scalar superfield $S(\hat{x},\bar{x},\theta,\bar{\theta})$ so that $\delta_{U}S= -i\hat{y}\hat{P}-i\bar{y}\bar{P}+\eta QS+\bar{\eta}\bar{Q}S$. From the Leibniz rule for the supercharges and the definition of covariant derivatives it follows that every polynomial in the superfields and its covariant derivatives is still a superfield.\newline\indent 

\subsection{Higher-momenta superalgebras}
The Lorentz-violating case admits a larger class of superalgebras compatible with the Coleman-Mandula theorem and the weighted power counting. The anticommutator $\lbrace Q_{\alpha},\bar{Q}_{\dot{\alpha}}\rbrace$ is in the $(1/2, 1/2)$ representation of the Poincar\'e algebra and has dimension $1$. In the Lorentz-invariant case this implies that it has to be proportional to $\sigma^{\mu}P_{\mu}$, which is the only  operator of dimension $1$ in the same representation.
In the Lorentz-violating case we can construct new operators $\bar{O}^{\bar\mu}$ of weight $[\bar{O}]=1$ and dimension $1$ in the vector representation of the $so(\bar{d})$ algebra. These operators are simply weighted polynomials of odd degree in the momentum $\bar{P}$ 
$$
\bar{O}^{\bar\mu}=\sum_{k}a_{k}\frac{\bar{P}^{2k}\bar{P}^{\bar\mu}}{\Lambda_{L}^{2k}}\ ,\ \ \ \text{with}\ \ \ k\leq\left[\frac{n-1}{2}\right]\ ,
$$
where the  constants $a_{k}$ weight $[a_{k}]=1-(2k+1)/n$ and $[a_{k}]\geq0$.  The parameter $n$ should be understood as the highest power of $\bar{P}$ that appears in the quadratic terms of the lagrangian, as explained in \cite{Anselmi:2007ri}. Therefore we can construct new superalgebras allowed by the Coleman-Mandula theorem and by the weighted power counting:
\begin{equation} 
\left\lbrace Q_{\alpha},\bar{Q}_{\dot{\alpha}}\right\rbrace=2\sigma^{\prime\mu}_{\alpha\dot{\alpha}}P_{\mu}+2\sigma^{\bar{\mu}}_{\alpha\dot{\alpha}}\bar{O}^{\bar{\mu}}\ \ \ ,\ \ \ \left\lbrace D_{\alpha},\bar{D}_{\dot{\alpha}}\right\rbrace=-2\sigma^{\prime\mu}_{\alpha\dot{\alpha}}P_{\mu}-2\sigma^{\bar{\mu}}_{\alpha\dot{\alpha}}\bar{O}^{\bar{\mu}}\ .\label{newalgebra} 
\end{equation}
The new operators $\bar{O}^{\bar\mu}$ have the same commutation rules as $\bar{P}^{\mu}$ in the SUSY algebra because, removing $J_{\hat{\mu}\bar{\nu}}$ from the original super-algebra $\mathfrak{s}\mathfrak{p}$, $\bar{P}^{2}$ becomes a Casimir operator for the new super-algebra $\mathfrak{s}\mathfrak{p}^{\prime}$.
If we take the low-energy limit $\Lambda_{L}\to \infty$ in (\ref{newalgebra}) we obtain again (\ref{eq.slv0}).
In principle $\mathfrak{s}\mathfrak{p}^{\prime}$ has the same superspace as $\mathfrak{s}\mathfrak{p}$ because we have not introduced new supercharges, but now our supercharges  will be non-linear operators in $\bar{P}$:
\begin{equation}
Q_{\alpha}=\frac{\partial}{\partial\theta^{\alpha}}-i\sigma^{\prime\mu}_{\alpha\dot{\alpha}}\bar{\theta}^{\dot{\alpha}}\partial_{\mu}-i\sigma^{\bar{\mu}}_{\alpha\dot{\alpha}}\bar{\theta}^{\dot{\alpha}}\bar{O}_{\bar{\mu}}\ , \label{supernew}
\end{equation}
so the operator $U(\hat{y},\bar{y}, \eta, \bar{\eta})=\text{e}^{i(\hat{y}\hat{P}+\bar{y}\bar{P}+i\eta Q+i\bar{\eta}\bar{Q})}$ is not a simple translation in superspace anymore. We can still define a superfield $S$ so that $\delta_{U}S= -i\hat{y}\hat{P}-i\bar{y}\bar{P}+\eta QS+\bar{\eta}\bar{Q}S$, but now it is in general not true that every polynomial in $S$ is still a superfield  because the supercharges (\ref{supernew}) do not respect Leibniz rule: $\delta_{U}(S_{1}S_{2})\neq\delta_{U}S_{1}S_{2}+S_{1}\delta_{U}S_{2}$. For this reason the superfield formalism loses its usefulness and it is not possible to promote directly the usual supersymmetric lagrangians for matter and gauge fields to lagrangians which are symmetric under the modified SUSY agebra (\ref{newalgebra}). The construction of a recent publication \cite{Xue:2010ih} is completely invalidated by this observation.
Actually, that construction works only for free theories because the Leibniz rule for the supercharges (\ref{supernew}) is fulfilled up to a total derivative if we consider only bilinears in the superfields. As an example of our general argument we consider the same theory as \cite{Xue:2010ih}, namely for (1,3) splitting we take the superalgebra (\ref{newalgebra}) for $n=3$. For the chiral supermultiplet we can write a kinetic lagrangian:
\begin{equation}\label{hd1}
\mathcal{L}_{kin}= \hat{\partial}\phi^{\dagger}\hat{\partial}\phi+(c_{1}\bar{\partial}+\frac{\bar{\partial}^{2}\bar{\partial}}{\Lambda_{L}^{2}})\phi^{\dagger}(c_{1}\bar{\partial}+\frac{\bar{\partial}^{2}\bar{\partial}}{\Lambda_{L}^{2}})\phi+\chi^{\dagger}(i\hat{\slashed\partial}+c_{1}i\bar{\slashed\partial}+\frac{i\bar{\slashed\partial}^{3}}{\Lambda_{L}^{2}})\chi+F^{\dagger}F \ .
\end{equation} 
The lagrangian is invariant under the transformations
\begin{align}\label{hd2}
&\delta_{\eta}\phi=\sqrt{2}\eta\chi\ ,\notag\\
&\delta_{\eta}\chi=i\sqrt{2}(\sigma^{\hat\mu}\bar{\eta}\partial_{\hat\mu}+\sigma^{\bar\mu}\bar{\eta}(c_{1}\partial_{\bar\mu}+\frac{\bar{\partial}^{2}\partial_{\bar\mu}}{\Lambda_{L}^{2}})) \phi+\sqrt{2}\eta F\ ,\notag\\
&\delta_{\eta}F=i\sqrt{2}\bar{\eta}(\bar{\sigma}^{\hat\mu}\partial_{\hat\mu}+c_{1}\bar{\sigma}^{\bar\mu}(c_{1}\partial_{\bar{\mu}}+\frac{\partial^{2}\partial_{\bar\mu}}{\Lambda_{L}^{2}}))\chi\ ,
\end{align}
which are clearly generated by supercharges that satisfy the superalgebra (\ref{newalgebra}) for $n=3$. The author \cite{Xue:2010ih} claims that if we want to add interactions invariant under (\ref{hd2}) to this theory it is sufficient to rephrase the usual superfield formalism for the new supercharges. Following his derivation, from the definition of the chiral superfield $\bar{D}_{\dot\alpha}\Phi=0$ we obtain:
\begin{equation}\begin{split}\label{hd3}
\mathop{\Phi(x,\theta)}= &\phi(x)+i\theta\sigma^{\hat{\mu}}\bar{\theta}\partial_{\hat\mu}\phi(x)-i\theta\sigma^{\bar\mu}\bar{\theta}(c_{1}\partial_{\bar\mu}+\frac{\bar\partial^{2}\partial_{\bar\mu}}{\Lambda_{L}^2})\phi(x)+\frac{1}{4}\theta^2\bar{\theta}^2(\hat{\partial}^2+2c_{1}\frac{\bar{\partial}^4}{\Lambda_{L}^2}+\frac{\bar{\partial}^{6}}{\Lambda_{L}^4})\phi(x)+\\
&+\sqrt{2}\theta\chi(x)+\frac{i}{\sqrt{2}}\theta^2\bar{\theta}(\bar{\sigma}^{\hat\mu}\partial_{\hat\mu}+\bar{\sigma}^{\bar\mu}(c_{1}\partial_{\bar\mu}+\frac{\bar{\partial}^{2}\partial_{\bar\mu}}{\Lambda_{L}^2}))\chi(x)+\theta^2 F(x)\ .
\end{split}
\end{equation}
If the superfield formalism worked, then each holomorphic function of the superfield (\ref{hd3}) would be invariant under (\ref{hd2}). As an example of a possible interaction lagrangian let us  consider  the usual cubic interaction in the superfields:
$$
\mathcal{L}_{int}=\frac{g}{3}\int d^{2}\theta\Phi^{3}+h.c= gF\phi^{2}-g\phi\psi\psi+h.c.\ .
$$
Now varying $\mathcal{L}_{int}$ with respect to (\ref{hd2}) we obtain 
\begin{equation*}\label{hd4}
\begin{split}
\delta\mathcal{L}_{int}= i\sqrt{2}\bar{\eta}[&\partial^{\hat\mu}(\bar{\sigma}_{\hat{\mu}}\chi\phi^2)+\partial^{\bar\mu}(c_{1}\bar\sigma_{\bar\mu}\chi\phi^2)+\frac{\partial^{\bar\mu}}{\Lambda_{L}^2}(\bar{\partial}^2\chi\phi^2-2\bar\sigma^{\bar\nu}\partial_{\bar\mu}\chi\partial_{\bar\nu}\phi\phi+2\bar{\sigma}^{\bar\nu}(\partial_{\bar\mu}\partial_{\bar\nu}\phi\phi+\partial_{\bar\mu}\phi\partial_{\bar\nu}\phi))] \\
&+i2\sqrt{2}\bar{\eta}\bar{\sigma}^{\bar\mu}\chi\partial^{\bar\nu}\partial_{\bar\mu}\phi\partial_{\bar\nu}\phi\ .\end{split}
\end{equation*}
Because of the last term in the previous equation $\delta\mathcal{L}_{int}\neq\partial^{\hat\mu}A_{\hat\mu}+\partial^{\bar\mu}B_{\bar\mu}$ and therefore the action is not invariant under the transformations (\ref{hd2}). This was expected, since the supercharges are non-linear in the spatial derivatives.\newline 
\indent Clearly, the possibility of constructing interacting quantum field theories which are invariant under the super-algebra (\ref{newalgebra}) is not ruled out by our observation and could be an interesting open field of research which, however, needs more involved constructions.\newline\indent
In the rest of this paper we will restrict ourselves to the case in which the supercharges are assumed to be linear in the spatial momenta and we will work with the superymmetric algebra (\ref{eq.slv0}).

\section{Renormalizable Lorentz-Violating Theories for Chiral Superfields} 
\subsection{General Discussion}
The integration measure in four dimensions for a general splitting $4=\hat{d}+\bar{d}$ weights $-\text{\dj}\equiv-(\hat{d}+\frac{\bar{d}}{n})$, so a renormalizable lagrangian has to be a weighted polynomial in the momenta of weight $\left[\mathcal{L}\right]=\text{\dj}$ and dimension $4$, as we want to keep the action dimensionless and weightless.\newline\indent 
Considering the super-algebra (\ref{eq.slv0}) we take a chiral superfield $\Phi(\hat{x},\bar{x},\theta,\bar{\theta})$ in the $N=1$ superspace defined by the constraint $\bar{D}\Phi=0$: 
\begin{equation*}
\Phi(\hat{x},\bar{x},\theta,\bar{\theta})=\phi+i\theta\sigma^{\prime\mu}\bar{\theta}\partial_{\mu}\phi+\frac{\theta^{2}\bar{\theta}^{2}}{4} \Box^{\prime} \phi+\sqrt{2}\theta\chi+\frac{i}{\sqrt{2}}\theta^{2}\bar\theta\bar{\sigma}^{\prime\mu}\partial_{\mu}\chi+\theta^{2} F\ .
\end{equation*}
The complex scalar field $\phi(x)$ and the Weyl spinor $\chi(x)$ are propagating quantum fields, and we can derive their weights from their kinetic lagrangian \cite{Anselmi:2007ri}. The chiral superfield then weights:
\begin{equation}
\left[\Phi\right]=\left[\phi\right]=\frac{\text{\dj}-2}{2}=\left[\theta\right]+\left[\chi\right]=\left[\theta\right]+\frac{\text{\dj}-1}{2}\ .\label{eq.slv.weight}
\end{equation}
The equation (\ref{eq.slv.weight}) fixes the weight of the spinor coordinates to $[\theta]=-1/2\  \forall\ n$, which is equal to the difference of weight  between scalars and fermions and is constant with respect to $n$. The spinor coordinates in the superspace do not rescale with $n$ and their weights are equal to their dimensions so that, as in the Lorentz invariant case, $\left[d^{4}\theta\right]=2$ and $\left[d^{2}\theta\right]=\left[d^{2}\bar{\theta}\right]=1$ $\forall\ n$. \newline\indent
We want to construct the most general Lorentz-violating lagrangian for $M$ different chiral superfields $\Phi_{i}$ with $i=1\dots M$. The K\"ahler potential $K[\Phi,\bar{\Phi}]$ weights $\left[K\right]=\text{\dj}-2$ and the superpotential $f[\Phi]$ weights $[f]=\text{\dj}-1$. If we demand polynomialilty of the lagrangian, $\left[\Phi\right]>0$, we obtain:
 \begin{equation}\label{eq.slv21}
\mathcal{L}_{(\hat{d},\bar{d})}=\int d^{4}\theta \bar{\Phi}_{i}\Phi_{i}+ \sum_{\alpha,N}\int d^{2}\theta\frac{\lambda_{N,\alpha}}{N\Lambda_{L}^{N(d-2)/2+p_{1}+p_{2}-3}}\left[\hat{\partial}^{p_{1}}\bar{\partial}^{p_{2}}\Phi^{N}\right]_{\alpha}+\text{h.c.}\ , 
\end{equation}
where $\alpha$ labels possible different derivative structures and a combinatorical factor can appear in the denominator if there are identical superfields. The most general K\"ahler potential which is renormalizable by weighted power counting has the same form as the Lorentz invariant one. This observation severely restricts the possibility of constructing supersymmetric Lorentz-violating models and at the same time will have important implications in the study of the RG flow at low energies of the $c$ parameter. In the superpotential, the derivative structure of the vertex defines a monomial in the superfield momenta of weight $\delta_{N}^{\alpha}=p_{1}+p_{2}/n$. If we want to preserve CPT invariance and symmetry under rotation in the submanifold $M_{\bar{d}}$ we have to assume that $p_{1}$ and $p_{2}$ are even numbers. The coupling constant $\lambda_{\alpha,N}$ associated to a vertex with $N$ superfields is a symmetric tensor with $N$ internal indices $i_{1}\dots i_{N}$, where $i_{k}=1\dots M\ \forall\ k$, and by power counting it has to weight
\begin{equation}\label{eq.slv23}
\left[\lambda_{N,\alpha}\right]=\text{\dj}-1-N\left[\Phi\right]-\delta_{N}^{\alpha}\ . 
\end{equation}
We will call a vertex weighted marginal when its coupling constant weights $\left[\lambda_{N,\alpha}\right]=0$, weighted relevant when $\left[\lambda_{N,\alpha}\right]>0$ and weighted irrelevant when $\left[\lambda_{N,\alpha}\right]<0$. As it has been shown in \cite{Anselmi:2007ri}, the renormalization rules in the Lorentz-violating case work as in the Lorentz invariant one after the substitution $d\rightarrow\text{\dj}$, so that we can express the renormalizabilty condition imposing that the weight of the coupling constant has to be greater or equal to zero, $\left[\lambda_{N,\alpha}\right]\geq0$. As we need $[\Phi]>0$ for polynomiality, taking $N=2$ in this inequality we can derive an upper bound on the weight of the monomial in the momenta $\delta_{N}^{\alpha}\leq1$, that ensures perturbative unitarity of the theory and forbids the presence of terms with time derivatives in the superpotential. We write $\mathcal{L}_{(\hat{d},\bar{d})}= \mathcal{L}_{kin}+ \mathcal{L}_{int}$, where the kinetic term of the lagrangian (\ref{eq.slv21}) is 
\begin{equation}\label{slv22}
\mathcal{L}_{kin}=\int d^{4}\theta\bar{\Phi}_{i}\Phi_{i}+\int d^{2}\theta\left(\sum_{l\leq[n/2]}\frac{(a_{l})_{ij}}{2\Lambda_{L}^{2l-1}} \bar{\partial}^{l}\Phi_{i}\bar{\partial}^{l}\Phi_{j}\right)+\text{h.c.}\ . 
\end{equation}
Only terms with an even number of space derivatives are allowed and the index $l$ in the sum is an integer that goes from zero to the integer part of the ratio $[n/2]$. The higher space derivatives terms generalize the mass term in the Wess-Zumino model \cite{Wess:1973kz} and are regulated by the coupling constants $(a_{l})_{ij}$, which are $M\times M$ symmetric matrix  of weight $[a_{l}]=1-2l/n$. The diagonal terms of $a_{l}$ behave as Majorana mass terms,  whereas the off-diagonal terms behave as Dirac mass terms. 
In order to simplify the notation we omit the internal indicex structure and take the free partition function of a theory with only one chiral superfield, in which the coupling constant $a_{l}$ becomes a coefficient and we can construct only Majorana mass terms. 
\begin{align*}
&Z_{0}\left[J,\bar{J}\right]=\int\mathcal{D}\Phi\mathcal{D}\bar{\Phi}\text{exp}-\left\lbrace \int d^{8}z\left[
\frac{1}{2}\begin{pmatrix}
                              \Phi&\bar{\Phi}
                        \end{pmatrix}A\begin{pmatrix}
							\Phi\\
							\bar{\Phi}
						   \end{pmatrix}-\begin{pmatrix}
									\Phi&\bar{\Phi}
								 \end{pmatrix}\begin{pmatrix}
										\frac{D^{2}}{4\Box^{\prime}}J\\
										\frac{\bar{D}^{2}}{4\Box^{\prime}}\bar{J}
										\end{pmatrix}\right]\right\rbrace \notag\\
&\ \text{and}\ A=\begin{pmatrix}
		A_{11}\frac{D^{2}}{4\Box^{\prime}} & 1\\
		1 & A_{22}\frac{\bar{D}^{2}}{4\Box^{\prime}}
	\end{pmatrix}\ \ \text{where}\ \ A_{11}=A_{22}=\left(\sum_{l\leq[n/2]}\frac{(-)^{l}a_{l}}{\Lambda_{L}^{2l-1}} \bar{\partial}^{2l}\right)\ .  
\end{align*} 
From the partition function we can derive the propagators for the chiral superfield using the methods of \cite{Grisaru:1979wc}:
\begin{align}
\left\langle\Phi(1)\bar{\Phi}(2)\right\rangle_{J=0}&=\frac{\delta_{12}}{p^{\prime 2}+\left(\sum_{l\leq[n/2]}\frac{a_{l}}{2\Lambda_{L}^{2l-1}}(\bar{p}^{2})^{l}\right)^{2}}\ ,\label{eq.slv5}\\
\left\langle\Phi(1)\Phi(2)\right\rangle_{J=0}&=\frac{D^{2}}{4}\left(\sum_{l\leq[n/2]}\frac{a_{l}(\bar{p}^{2})^{l}}{\Lambda_{L}^{2l-1}}\right)\frac{\delta_{12}}{p^{\prime2}\left(p^{\prime2}+\left(\sum_{l\leq[n/2]}\frac{a_{l}(\bar{p}^{2})^{l}}{\Lambda_{L}^{2l-1}}\right)^{2}\right)}\ ,\label{eq.slv6}
\end{align}
where we have reabsorbed the $-\frac{D^{2}}{4}$ factors in the Feynman rules for the vertices and $p^{\prime2}$ is defined in (\ref{pprimo}).  If we differentiate the propagators (\ref{eq.slv5}) and (\ref{eq.slv6}) with respect to any coefficient $a_{l}$ with $l<[n/2]$ the weight of the denominator increases by $1-2l/n$ and differentiating with respect to $c$  increases by $1-1/n$. Hence we can make any Feynman graph convergent by differentiating it a suitable number of times with respect to the coefficients $a_{l}$ or $c$, and the counterterms will be polynomials in $a_{l}$ and $c$. We can consider the super-renormalizable operators associated with the coefficients $a_{l}$ and $c$ as vertices with two external lines and treat them perturbatively. Doing that, we can study the UV behavior of the Lorentz-violating theories keeping in the propagators (\ref{eq.slv5}) and (\ref{eq.slv6}) only the terms with the maximum number of spatial derivatives:
\begin{align}
\mathcal{P}_{\Phi\bar{\Phi}}&=\frac{\delta_{12}}{\hat{p}^{2}+a^{2}_{[n/2]}\frac{(\bar{p}^{2})^{2[n/2]}}{\Lambda_{L}^{4[n/2]-2}}}\ ,\label{eq.slv7}\\ 
\mathcal{P}_{\Phi\Phi}&=\frac{D^{2}}{4}\frac{a_{[n/2]}(\bar{p}^{2})^{[n/2]}}{\Lambda_{L}^{2[n/2]-1}}\frac{\delta_{12}}{p^{\prime2}\left(\hat{p}^{2}+a^{2}_{[n/2]}\frac{(\bar{p}^{2})^{2[n/2]}}{\Lambda_{L}^{4[n/2]-2}}\right)}\ .\label{eq.slv8}
\end{align}
The weight of the coefficient $a_{[n/2]}$ is zero for even $n$ and strictly positive for odd $n$, and in fact the operators $\left(\bar{\partial}^{[n/2]}\Phi\right)^{2}$ in the free lagrangian (\ref{slv22}) are strictly renormalizable for even $n$ and super-renormalizable for odd $n$. Therefore, for odd $n$ we cannot construct propagators which are the inverse of homogeneous polynomials of weight 2 and this fact completely invalidate our construction. To understand what does not work in the odd $n$ case we compare the kinetic terms of the fermionic and the scalar lagranians in the non-supersymmetric case for an arbitrary $n$ \cite{Anselmi:2007ri}:
\begin{align}
&\mathcal{L}_{s}=\frac{1}{2}(\hat{\partial}\phi)^2-\frac{c^2}{2}(\bar\partial\phi)^2-\sum_{l=2}^{n}\frac{a_{l}^{2}}{2\Lambda_{L}^{2l-2}}(\partial^{l}\phi)^2-\frac{m^{2}}{2}\phi^2 \label{fine1}\ ,\\
&\mathcal{L}_{f}=\bar{\psi}(i\hat{\slashed\partial}+vi\bar{\slashed\partial}+\sum_{l\text{ odd}}^{n}\frac{b^{\prime}_{l}}{\Lambda_{L}^{l-1}}(i\bar{\slashed\partial})^{l}+\sum_{l\text{ even}}^{n}\frac{b_{l}}{\Lambda_{L}^{l-1}}(i\bar{\slashed\partial})^{l}-M)\psi\ .\label{fine2}
\end{align}
From the equations of motion associated with the lagrangians (\ref{fine1}) and (\ref{fine2}) we derive the corresponding dispersion relations\footnote{For simplicity we write the dispersion relations in the case of $(1,3)$ splitting, in which there is a natural identification of the energy with the only component of the four momentum whose weighted dimension coincides with the usual one. The extension to different splittings is straightforward.}: 
\begin{align}
&E^2_{s}(\bar{p})= c^2\bar{p}^{2}+\sum_{l=2}^{n}a^{2}_{l}\frac{\bar{p}^{2l}}{\Lambda_{L}^{2l-2}}+m^2\label{fine3}\ ,\\   
&E^2_{f}(\bar{p})= \bar{p}^{2}(v+\sum_{l\text{ odd}}^{n}\frac{b_{l}^{\prime}}{\Lambda_{L}^{l-1}}\bar{p}^{l-1})^2+(M+\sum_{l\text{ even}}^{n}\frac{b_{l}}{\Lambda_{L}^{l-1}}\bar{p}^{l})^2\ .\label{fine4}
\end{align}
In the Lorentz invariant case the dispersion relation among energy and spatial momentum is universal for all particles $E^2(\bar{p})=c^2\bar{p}^{2}+m^{2}$.  Conversely, we see from (\ref{fine4}) that in the Lorentz-violating  case the dispersion relation for fermions contains two different kind of contributions that are related respectively to terms with an even or an odd number of derivatives in the kinetic lagrangian (\ref{fine2}). This happens because the terms with an even number of derivatives behave like mass terms from the point of view of the spin $1/2$ representation of the Lorentz group, while the terms with an odd number of derivatives behave like $\bar{\slashed{p}}$. A necessary condition for the theory to be supersymmetric is that all the dispersion relations of particles in the same supermultiplet have to be the same. We have shown that all the higher spatial derivatives terms that behave like masses can be supersymmetrized by adding appropriate F-terms to the superpotential. On the contrary, all terms with an odd number of higher spatial derivatives would correspond to modifications of the K\"alher potential. However, a K\"ahler potential which is renormalizable by weighted power counting should have the same form as the Lorentz invariant one and therefore we cannot construct a supersymmetric version of a theory with an odd number of higher spatial derivatives in the fermionic kinetic term. If $n$ is odd this means that it is not possible to construct a supersymmetric version of a free theory with scalars and fermions. For even $n$ we can construct supersymmetric theories in which dispersion relations for fermions will be of the form (\ref{fine4}) with $b^{\prime}_{l}=0$ $\forall\ l$. \newline\indent   
The Feynman rules for the vertices remain unchanged with respect to the Lorentz invariant case \cite{Grisaru:1979wc} because the Lorentz-violating terms do not modify the $\theta$-structure of Feynman graphs. Therefore the divergent contributions to the effective energy are polynomials in the external momenta of the form
\begin{equation}\label{eq.slv99}
\Gamma_{\infty}=\int d^{4}xd^{4}\theta F(\Phi,\bar{\Phi},D\Phi\dots,\bar{\partial}\Phi,\dots)\ . 
\end{equation}
By power counting we obtain $[F]=\text{\dj}-2$, so that the non-renormalization theorem \cite{Gates:1983nr} for the divergent contributions still works in the Lorentz-violating case and the divergent contributions affect only the K\"ahler potential.
We can calculate the superficial divergence for a generic Feynman graph $G$ at $L$ loops, with $I$ propagators, $V$ vertices and $E$ external lines\footnote{The covariant derivatives algebra contains the positive weighted constant $c$, so that their commutation rules generate non-homogeneous polynomials of degree $1$ in the momenta. A chiral vertex $\int d^{2}\theta\Phi^{N}$ yelds $N-1$ $D^{2}$ factors to the numerators, that correspond to a non-homogeneous polynomial of maximum degree $N-1$ in the momenta. In the computation of the superficial divergence we consider the term of maximum degree for every polynomial generated by the covariant derivatives commutation rules. The other terms will have a better UV behavior.} 
\begin{equation}
\begin{split}
\omega(G)&=(\text{\dj}-2)L-2I-E+\sum_{N,\alpha}v_{N}\left(N-1+\delta^{\alpha}_{N}\right)\\
	 &=d(E)-2-\sum_{N,\alpha} v_{N}^{\alpha}[\lambda_{\alpha,N}]\ ,\ \text{where}\ d(E)= \text{\dj}-E[\Phi]
\end{split} 
\end{equation}
and the weights of the chiral superfield $\Phi$ and of the coupling constant are defined in (\ref{eq.slv.weight}) and (\ref{eq.slv23}). For renormalizable theories, taking $E=2$ yields an upper bound for the superficial divergence, $\omega(G)\leq0$, that is in agreement with the result of the non-renormalization theorem (\ref{eq.slv99}). Therefore in a supersymmetric Lorentz-violating theory the K\"ahler potential can receive radiative corrections with logarithmic divergences if and only if there are strictly renormalizable interactions. If the theory contains only super-renormalizable interactions, then the theory is finite.
In the Lorentz invariant Wess-Zumino model, the non-renormalization theorem ensures that the behavior of the theory at different energies is regulated only by  the wave function renormalization constant. In our models this is not true anymore. However, we can derive  relations among different renormalization constants. The most general renormalizable interaction lagrangian contains two different kinds of composite operators:
\begin{equation*}
\mathcal{L}_{int}\supset \int d^{2}\theta \left\lbrace\frac{\lambda_{k}}{\Lambda_{L}^{k-3}}\Phi^{k}+\frac{\lambda^{\prime}_{k,l,\alpha}}{\Lambda_{L}^{k+2l-3}}\left[\bar{\partial}^{2l}\Phi^{k}\right]_{\alpha}\right\rbrace+ \text{h.c.}\ .  
\end{equation*}
Demanding the renormalizability of the theory it is easy to see that $k$ is an integer number $2<k\leq\bar{N}$, where $\bar{N}$ is the maximum number of legs for the chiral vertex in a Lorentz-violating theory with fixed $n$,
\begin{equation}\label{eq.slv11}
\bar{N}=\left[2\frac{\text{\dj}-1}{\text{\dj}-2}\right]\ ,
\end{equation}
and we indicate with the squared brackets the integer part of the enclosed ratio.\newline\indent 
Consequently $n/2\leq l<n$, where in our case $n/2$ is alway an integer because our theories are defined for even $n$ only. We will classify all the possible models for different splittings in section (\ref{section5}).
The renormalization constant of operators like $\Phi^{k}$ is simply the product of $k$ times the renormalization constant of the superfield $\Phi$, whereas composite operators like $O_{k,l,\alpha}\equiv\left[\bar{\partial}^{2l}\Phi^{k}\right]_{\alpha}$ renormalize in a non-trivial way because of the space derivatives in the vertices.
Rewriting the lagrangian respect to the \textit{bare} quantities we get
\begin{align*}
&(\lambda_{k})_{B}=\mu^{\epsilon(k/2-1)}Z_{\lambda_{k}}\lambda_{k}\ , \ \Phi_{B}=Z_{\Phi}^{1/2}\Phi\ ,\ (a_{l})_{B}=Z_{a_{l}}a_{l}\ ,\ c_{B}= c+\Delta_{c}\ ,\\
&(\lambda^{\prime}_{k,l,\alpha})_{B}=\mu^{\epsilon(k/2-1)}Z_{\lambda^{\prime}_{k,l,\alpha}}\lambda^{\prime}_{k,l,\alpha}\ ,\ (O_{k,l,\alpha})_{B}=Z_{O_{k,l,\alpha}}O_{k,l,\alpha}\ ,\ \Lambda_{LB=}Z_{\Lambda_{L}}\Lambda_{L}\ .
\end{align*}
From the non-renormalization of the superpotential we then obtain:
\begin{equation}
\frac{Z_{\Phi}^{k/2}Z_{\lambda_{k}}}{Z_{\Lambda_{L}}^{k-3}}=1\ \ ,\ \ \frac{Z_{O_{k,l,\alpha}}Z_{\lambda^{\prime}_{k,l,\alpha}}}{Z_{\Lambda_{L}}^{k+2l-3}}=1\ .
\end{equation}
On the other hand the K\"ahler potential is renormalized by a single superfield wave function renormalization $K_{B}[\bar\Phi,\Phi]=Z_{\Phi}K[\bar\Phi,\Phi]$, so that:
\begin{equation}
\Delta_{c}=0\ \ ,\ \ \frac{Z_{a_{l}}}{Z_{\Lambda_{L}}}=1\ .
\end{equation}
This last result implies that the deviation from the speed of light of a supermultiplet does not renormalize in a supersymmetric theory. We will discuss in more detail the physical consequences of this result in section (\ref{section6}).

\subsection{Homogeneous Theories} 
\indent  
We call homogeneous a theory whose vertices are all marginal. This kind of theories is invariant at the classical level under the weighted dilatations (\ref{1}), while at the quantum level the weighted scale invariance is anomalous and related to the renormalization group \cite{Anselmi:2007ri}. Moreover in the non-supersymmetric case homogeneous models for scalars and spinors were classified in \cite{Anselmi:2007ri}. In this section we want to study the possibility of constructing homogeneous Lorentz-violating supersymmetric theories. To analyze this problem it is important to emphasize that in the supercharges algebra (\ref{eq.slv0}) Lorentz violation is introduced by means of the weighted constant $c$. Therefore if $[c]>0$ and $c\neq0$ the kinetic term in the K\"ahler potential always introduces a non-homogeneous term in the propagator, which breaks the weighted scale invariance. \newline\indent We can define Lorentz-violating homogeneous theories if $[c]=0$ and $n=1$. In this case we obtain the model proposed in \cite{Berger:2001rm}, in which the interaction sector is equal to the Wess-Zumino model.
However, we can obtain another class of homogeneous theories by taking $c=0$. In this case the supercharges algebra (\ref{eq.slv0}) becomes:
\begin{equation}
\left\lbrace Q_{\alpha},\bar{Q}_{\dot{\alpha}}\right\rbrace= 2\sigma^{\hat{\mu}}_{\alpha\dot{\alpha}}P_{\hat{\mu}}\ . \label{mucchi}
\end{equation}
The resulting algebra (\ref{mucchi}) is the usual $N=1$ supersymmetry algebra in $d=4$, but projected on the submanifold $M_{\hat{d}}$. It is clear that the K\"ahler potential for the chiral superfields associated with this algebra does not introduce non-homogeneous terms in the propagators. Therefore we can define a class of free homogeneous lagrangians for every even value of $n$, inserting in the superpotential (\ref{slv22}) only the bilinear with the maximum number of spatial derivatives, regulated by the weightless constant $a_{n/2}$. The propagators are the inverse of homogeneous polynomials of weight $2$:
\begin{align}\label{propmucchi}
& \langle\Phi(1)\bar{\Phi}(2)\rangle=\frac{\delta_{12}}{\hat{p}^{2}+a^{2}_{n/2}\frac{\bar{p}^{2n}}{\Lambda_{L}^{2n-2}}}\ ,\notag \\
& \langle\Phi(1)\Phi(2)\rangle=\frac{D^{2}}{4}\frac{a_{n/2}(\bar{p}^{2})^{n/2}}{\Lambda_{L}^{n-1}}\frac{\delta_{12}}{\hat{p}^{2}\left(\hat{p}^{2}+a^{2}_{n/2}\frac{(\bar{p}^{2})^{n}}{\Lambda_{L}^{2n-2}}\right)}\ .
\end{align}
From these free theories we can construct interacting lagrangians by adding all the renormalizable terms in the superpotential. When $c=0$, if we consider only the strictly renormalizable interactions in the superpotential, we then obtain homogenous interacting theories. At the quantum level the fixed points of the renormalization group for these theories are still invariant under weighted scale transformations, but far from the fixed points the symmetry is anomalous.
In the low-energy limit we cannot restore the usual  $N=1$ supersymmetry algebra in $d=4$ and we cannot obtain a Lorentz-invariant theory with usual propagators because of the weighted scale invariance; in fact super-renormalizable terms cannot be produced by renormalization because they would break the weighetd scale invariance. In the IR limit the propagators (\ref{propmucchi}) become
\begin{equation}
\lim_{\Lambda_{L}\to\infty}\langle\Phi(1)\bar{\Phi}(2)\rangle= \frac{\delta_{12}}{\hat{p}^{2}}\ \ \ \ ,\ \ \ \ \lim_{\Lambda_{L}\to\infty}\langle\Phi(1)\Phi(2)\rangle =0\label{proplim}\ .
\end{equation}
The propagators (\ref{proplim}) do not depend on $\bar{p}$, so that all the diagrams constructed with these propagators are not computable, because they contain divergences that no counterterms can eliminate. Hence the IR limit is singular. From homogenous theories we can construct non-homogeneous theories invariant under the algebra (\ref{mucchi}), by adding to the superpotential all the super-renormalizable terms. Doing that, we are breaking the weighted scale invariance, but terms fundamental for the Lorentz symmetry recovery such as $\phi^{\ast}\partial^{\bar{\mu}}\partial_{\bar{\mu}}\phi$ or $i\partial_{\bar{\mu}}\bar{\psi}\bar{\sigma}^{\bar{\mu}}\psi$ are not generated by renormalization because they break the symmetry of the lagrangian under supersymmetry transformations (\ref{mucchi}). This means that for $c=0$ Lorentz symmetry cannot be restored and the IR limit of these theories is still singular.

\subsection{Classification of neutral chiral superfield's models}\label{section5}

We want to classify all the possible theories invariant under the superalgebra (\ref{eq.slv0}) for all possible splittings of the four dimensional spacetime manifold. 
The basic ingredient of such classification will be the maximum number of legs for a chiral vertex defined in (\ref{eq.slv11}). 
\begin{enumerate}
 \item For splitting (0,4) we have $\bar{N}=\left[2\frac{4-n}{4-2n}\right]\ .$ The only solution is  the $n=1$ and $\bar{N}=3$ and we thus recover the  Wess-Zumino model. 
\item For (1,3) splitting we obtain $\bar{N}=\left[2\frac{3}{3-n}\right]$. For $n=1$ we find again the Lorentz-violating Wess-Zumino model proposed in \cite{Berger:2001rm, Berger:2003ay}. Taking $n=2$ we find $\bar{N}=6$, which is the only non-trivial solution of the condition (\ref{eq.slv11}). Hence, for $(1,3)$ splitting the only theory with strictly renormalizable interactions has $n=2$ and $\text{\dj}=5/2$. The Lagrangian, requiring symmetry under the transformation $\Phi\rightarrow-\Phi$, will be
\begin{equation}
\mathcal{L}=\int d^{4}\theta\bar{\Phi}\Phi+\int d^{2}\theta\left[ \frac{(\bar{\partial}\Phi)^{2}}{2\Lambda_{L}}+\frac{m\Phi^{2}}{2}\right]+\int d^{2}\theta\left[\lambda_{4}\frac{\Phi^{4}}{4!\Lambda_{L}}+ \lambda_{6}\frac{\Phi^{6}}{6!\Lambda_{L}^{3}}\right]+\text{h.c.}\ . 
\end{equation}
We can derive the propagators for the superfields and expand them for high momenta in order to study the UV behavior of the theory:
\begin{align}
&\mathcal{P}_{\Phi\bar{\Phi}}= \frac{\delta_{12}}{\hat{p}^{2}+(c^{2}+\frac{2m}{\Lambda_{L}})\bar{p}^{2}+\frac{\bar{p}^{4}}{\Lambda_{L}^{2}}+m^{2}}\simeq\frac{\delta_{12}}{\hat{p}^{2}+\frac{\bar{p}^{4}}{\Lambda_{L}^{2}}}\label{prop1}\ ,\\
&\mathcal{P}_{\Phi\Phi}=\frac{D^{2}}{4} \frac{\bar{p}^{2}}{\Lambda_{L}} \frac{\delta_{12}}{\hat{p}^{2}+(c^{2}+\frac{2m}{\Lambda_{L}})\bar{p}^{2}+\frac{\bar{p}^{4}}{\Lambda_{L}^{2}}+m^{2}}\simeq \frac{D^{2}}{4} \frac{\bar{p}^{2}}{\Lambda_{L}}\frac{\delta_{12}}{p^{\prime 2}\left(\hat{p}^{2}+\frac{\bar{p}^{4}}{\Lambda_{L}^{2}}\right)}\label{prop2}\ .
\end{align}
The first divergent radiative correction to the K\"ahler potential is at  $4$ loops:
$$
\raisebox{-20mm}{\begin{picture}(120,100)
   \Line(0,60)(30,60)
   \Text(10,65)[lb]{\Black{$\Phi$}}
    \Arc(60,60)(40,0,360)
    \Oval(60,60)(20,40)(0)
    \Line(30,60)(90,60)
    \Line(90,60)(120,60)
    \Text(110,65)[lb]{\Black{$\bar{\Phi}$}}
    \Vertex(100,60){2}  
  \end{picture}}= \frac{\lambda_{6}^{2}}{5!}\int \frac{d\hat{k}}{2\pi}\frac{d^{3}\bar{k}}{(2\pi)^{3}}\int d^{4}\theta_{1}d^{4}\theta_{2}\Phi(-k,\theta_{1})\Phi(k,\theta_{2})\mathcal{B}\mathcal{D}\ .
$$
The covariant derivatives algebra is easy to compute if we recall the usual identities
$$
\mathcal{D}= \delta_{12}\frac{D^{2}\bar{D}^{2}}{16}\delta_{12}\frac{\bar{D}^{2}D^{2}}{16}\delta_{12}\frac{D^{2}\bar{D}^{2}}{16}\delta_{12}\frac{D^{2}\bar{D}^{2}}{16}\delta_{12}=\delta_{12} \ .
$$
Therefore, the superfields computation is reduced to the computation of the bosonic integral $\mathcal{B}$, which is not easily computable because of the modified form of the propagators (\ref{prop1}) and (\ref{prop2}).
\item For $(2,2)$ splitting $\bar{N}=n+2$ and we can construct theories with strictly renormalizable interactions for any even $n$. For example we can choose $n=2$ and write the complete theory:
\begin{equation*}
\mathcal{L}=\int d^{4}\theta \bar{\Phi}\Phi+\int d^{2}\theta\left[\frac{(\bar{\partial}\Phi)^{2}}{2\Lambda_{L}}+ \frac{m}{2}\Phi^{2}\right]+\int d^{2}\theta \left[ \frac{\lambda_{3}}{3!}\Phi^{3}+\frac{\lambda_{4}}{4!\Lambda_{L}}\Phi^{4}\right]\ . 
\end{equation*}
The kinetic term is the same of the theory $n=2$ for $(1,3)$ splitting and the propagators are (\ref{prop1}) and (\ref{prop2}).\newline The first divergent radiative correction to the K\"ahler potential is at $2$ loops:
$$
\raisebox{-13mm}{\begin{picture}(100,80)
   \Line(0,40)(20,40)
   \Text(10,45)[lb]{\Black{$\Phi$}}
    \Arc(50,40)(30,0,360)
    \Line(20,40)(80,40)
    \Line(80,40)(100,40)
    \Text(90,45)[lb]{\Black{$\bar{\Phi}$}}
    \Vertex(80,40){2}  
  \end{picture}}= \frac{\lambda_{4}^{2}}{3!}\int \frac{d^{2}\hat{k}}{(2\pi)^{2}}\frac{d^{2}\bar{k}}{(2\pi)^{2}}\int d^{4}\theta_{1}d^{4}\theta_{2}\Phi(-k,\theta_{1})\Phi(k,\theta_{2})\mathcal{B}\mathcal{D}
$$
Again, for this graph $\mathcal{D}=\delta_{12}$ but the bosonic integral $\mathcal{B}$ is again very hard to compute.
\item In the $(3,1)$ case we obtain $\bar{N}=\left[2\frac{2n+1}{n+1}\right]$ that has only one integer solution for $n=1$ that is the trivial one. Therefore we can construct only super-renormalizable theories.
\item For splitting $(4,0)$ we obtain the Lorentz invariant Wess-Zumino model.
\end{enumerate}
\section{Gauge invariant theories}  
We want to study the problem of finding a gauge invariant version of the Lorentz-violating supersymmetric theories that we have found in section 3. First of all we apply a result of  \cite{GrootNibbelink:2004za} to show that it is not possible to generalize the theories of section 3 to the case of charged chiral superfields. 
The basic observation in our construction was that it is possible to insert higher space derivatives as mass terms in the superpotential in (\ref{slv22}) because they preserve the chirality of $\Phi$, which is clear recalling that $[D_{\alpha},\partial_{\bar{\mu}}]=0$.
Charged chiral superfields transform with a phase under the action of a general $SU(N)$ gauge group
\begin{equation*}
\Phi_{i}\rightarrow e^{\Lambda}\Phi_{i}\ , 
\end{equation*}
where $\Lambda$ is a chiral superfield. We can define a gauge invariant version of the supersymmetric covariant derivative:
\begin{equation*}
\mathcal{D}_{\alpha}\Phi_{i}=e^{-V}D_{\alpha}(e^{V}\Phi_{i})\ ,
\end{equation*}
where $V(\hat{x}, \bar{x}, \theta, \bar{\theta})$ is the vector superfield with its usual gauge trasformation law: $e^{V}\rightarrow e^{\bar{\Lambda}}e^{V}e^{-\Lambda}$. It is clear that explicit spatial derivatives in the action break gauge invariance. In principle, as was observed in \cite{GrootNibbelink:2004za}, we could still introduce higher spatial derivatives in the superpotential promoting $\partial_{\bar{\mu}}$ to a covariant derivative:
\begin{equation}
D_{\bar{\mu}}\Phi=-\frac{i\bar{\sigma}_{\bar{\mu}}^{\dot{\alpha}\alpha}}{4}\bar{D}_{\dot{\alpha}}\mathcal{D}_{\alpha}\Phi\ .
\end{equation}
The problem, however, is that $D_{\bar{\mu}}$ does not preserve the chirality condition. In fact, as it was checked in \cite{GrootNibbelink:2004za}:
\begin{equation}
\bar{D}_{\dot\alpha}\bar{D}_{\dot\beta}\left(e^{-V}D_{\alpha}e^{V}\Phi_{i}\right)=2\varepsilon_{\dot{\beta}\dot{\alpha}}W_{\alpha}\Phi\neq0\ . 
\end{equation}
This argument shows that the theories that we have constructed are not generalizable to the case of charged chiral superfields.\newline\indent
Now we will see directly in the gauge sector that, requiring gauge invariance and supersymmetry, the weighted power counting has to coincide with the usual one. We briefly review the derivation of the weights for the gauge fields referring to \cite{Anselmi:2008bq, Anselmi:2008bs} for a complete study of Lorentz-violating gauge theories. In the Lorentz-violating case the gauge field $A_{\mu}$ has to be decomposed as all the other vectors into time and space components $A=(\hat{A}, \bar{A})$.  Therefore the covariant derivative is decomposed as 
\begin{equation}
D=(\hat{D},\bar{D})=(\hat{\partial}+e\hat{A}, \bar{\partial}+e\bar{A})\ ,\label{covariant}
\end{equation}
where $e$ is the gauge coupling constant. To be consistent with the definition of covariant derivative (\ref{covariant}) we have to weight $[e\hat{A}^{\prime}]=[\hat{\partial}]=1$ and $[e\bar{A}]=[\bar{\partial}]=1/n$. The field strength is split into three parts:
\begin{equation*}
\hat{F}_{\mu\nu}\equiv F_{\hat{\mu}\hat{\nu}}\ ,\ \ \ \ \tilde{F}_{\mu\nu}\equiv F_{\hat{\mu}\bar{\nu}}\ , \ \ \ \ \bar{F}_{\mu\nu}\equiv F_{\bar{\mu}\bar{\nu}}\ . 
\end{equation*}
The kinetic lagrangian has to contain $(\hat{\partial}\hat{A})^{2}$, so we can obtain the weight of $\hat{A}$, that is equal to the weight of the scalar field, and from the definition of covariant derivatives we derive the weights for $\bar{A}$ and $[e]=2-\text{\dj}/2$. Summarizing we have:
\begin{equation}\label{ass1}
[\hat{A}]=\frac{\text{\dj}-2}{2}\ ,\ \ [\bar{A}]=\frac{\text{\dj}}{2}-2+\frac{1}{n}\ ,\ \ [\hat{F}]=\frac{\text{\dj}}{2}\ ,\ \ [\tilde{F}]=\frac{\text{\dj}}{2}-1+\frac{1}{n}\ ,\ \ [\bar{F}]=\frac{\text{\dj}}{2}-2+\frac{2}{n}\ .
\end{equation}  
The requirement of absence of spurious subdivergences \cite{Anselmi:2008bq} implies that 
$$
\hat{d}=1\ ,\ \ \text{\dj}<2+\frac{2}{n}\ ,\ \ d=\text{even}\ ,\ \ n=\text{odd}\ ,
$$ 
and the only acceptable splitting is $(1,3)$. In this case we have $\hat{F}=0$ so it is possible to rearrange the weights of the gauge field and the gauge coupling so that the product $eA$ maintains the same weight and at the same time $[\tilde{F}]=\text{\dj}/2$:
\begin{equation}\label{ass2}
[\hat{A}]=\frac{\text{\dj}}{2}-\frac{1}{n}\ ,\ \ [\bar{A}]=\frac{\text{\dj}}{2}-1\ ,\ \ [\tilde{F}]=\frac{\text{\dj}}{2}\ ,\ \ [\bar{F}]=\frac{\text{\dj}}{2}-1+\frac{1}{n}\ ,\ \ [e]=1+\frac{1}{n}-\frac{\text{\dj}}{2}\ . 
\end{equation}
In the supersymmetric case both weight assignments (\ref{ass1}) and (\ref{ass2}) have to be consistent with the relations among the weights of the fields imposed by the supersymmetric transformations generated by the supercharges (\ref{eq.slv0}). For the vector multiplet the supersymmetric transformation are:
\begin{align}\label{transf}
&\delta_{\eta}A^{\mu}=\bar{\eta}\bar{\sigma}^{\mu\prime}\lambda+\bar{\lambda}\bar{\sigma}^{\mu\prime}\eta\ ,\notag\\
&\delta_{\eta}\lambda=\frac{i}{2}\sigma^{\mu\prime}\bar{\sigma}^{\nu\prime}\eta F_{\mu\nu}+\eta D\ ,\notag\\
&\delta_{\eta}D=\bar{\eta}\bar{\sigma}^{\mu\prime}\partial_{\mu}\lambda-\partial_{\mu}\bar{\lambda}\bar{\sigma}^{\mu\prime}\eta\ ,
\end{align} 
where the gaugino $\lambda$ is a propagating Majorana fermion, $D$ is an auxiliary field and $\eta$ is the spinorial parameter of the supersymmetry transformation. Since we know the weights of $\lambda$, $\eta$ and of the weighted constant $c$, we can obtain the weights of the other fields of the supermultiplet, applying the weighted power counting to the relations (\ref{transf}) yelds
\begin{align}
&[\hat{A}]=\frac{\text{\dj}-2}{2}\ ,\ \ [\bar{A}]=\frac{\text{\dj}}{2}-\frac{1}{n}\ ,\ \ [D]=\frac{\text{\dj}}{2}\ ,\notag\\
&[\hat{F}]=\frac{\text{\dj}}{2}\ ,\ \ [\tilde{F}]=\frac{\text{\dj}}{2}-(1-\frac{1}{n})\ ,\ \ [\bar{F}]=\frac{\text{\dj}}{2}-2(1-\frac{1}{n})\ .\label{susyweight}
\end{align}
Any gauge theory that has a supersymmetric extension invariant under the supersymmetry algebra (\ref{eq.slv0}) has to satisfy the constraint on the weight of the fields (\ref{susyweight}). Therefore, we can take the two possible weight assignments for Lorentz-violating gauge field theories (\ref{ass1}) and (\ref{ass2}) and check for which value of $n$ these theories can admit a supersymmetric extension.
Doing that we found that the only possible value is $n=1$ in both cases and hence, as long as we require supersymmetry and gauge invariance, we have to weight time and space in the same way, regardless of the condition of absence of spurious divergences.  The only Lorentz-violating operators are introduced by the weighted constant $c$. These operators are renormalizable in the usual sense and correspond to the CPT preserving operator in the gauge sector of the SME \cite{Colladay:1998fq}. In particular they can be expressed introducing a twisted derivative $\tilde{\partial}_{\mu}=\partial_{\mu}+k_{\mu}^{\nu}\partial_{\nu}$ \cite{Colladay:2010xf}, where in our case $k_{\mu\nu}=(c-1)\delta_{\bar{\mu}\bar{\nu}}$. 
We will show in the next section that the constant $c$ does not renormalize in the supersymmetric case. As we need to fine tune $c$ in order to recover Lorentz symmetry  at low energies, the Lorentz-violating parameters will be extremely small also at high energies. This argument shows that demanding renormalizability and gauge invariance for supersymmetric theories, the Lorentz invariance follows as a consequence.

\section{Low-energy limit and Lorentz symmetry recovery}\label{section6}

If we consider Lorentz-violating theories as candidates to describe high-energy physics, then Lorentz invariant theories are effective field theories which describe physics at low energies with respect to $\Lambda_{L}$. In the framework of renormalizable Lorentz-violating theories it is still true that  the renormalization procedure commutes with the low-energy limit. Therefore it is a general result that a Lorentz-violating high-energy theory renormalizable by weighted power counting tend to a low-energy theory which is renormalizable by usual power counting and contains Lorentz-violating parameters. The low-energy theory is then obtained from the high-energy one simply by taking the limit $\Lambda_{L}\to\infty$. The recovery of Lorentz invariance at low energies is regulated by the RG behavior of the Lorentz-violating parameters at low energies that correspond to operators of dimension less than or equal to four, which are not suppressed by any power of $\Lambda_{L}$. \newline
\indent In this section we want to study how the Lorentz recovery problem at low energies is modified in presence of supersymmetry.  
If supersymmetry is an exact symmetry of nature at high energies  then in the low-energy limit supersymmetry has to be broken for several phenomenological reasons \cite{Martin:1997ns}. Ignoring the exact mechanism of SUSY breaking we can parametrize the supersymmetry breaking at low energies by adding explicit breaking terms to the supersymmetric lagrangian. We require the breaking to be soft, in the sense that the supersymmetry breaking terms should not generate quadratic divergences. 
It has been pointed out in \cite{Girardello:1981wz} that the natural setting for studying the low-energy limit of supersymmetric theories in presence of soft breaking terms is the superfield formalism. The soft breaking terms are parametrized as couplings among dynamical superfields and external spurion superfields. The possible types of spurion superfields which break supersymmetry softly for Lorentz invariant theories are classified in \cite{Girardello:1981wz}.\newline
\indent On this basis we can compute the low-energy limit for the general supersymmetric Lorentz-violating lagrangian (\ref{eq.slv21}). Assuming that the supersymmetry breaking soft terms are generated at energy $\mu_{s}<< \Lambda_{L}$ we obtain the standard Lorentz-invariant soft terms: 
\begin{equation}\begin{split}
\mathcal{L}=& \int d^{4}\theta \bar{\Phi}_{i}\Phi_{i}+ \int d^{2}\theta ( \frac{m_{ij}}{2}\Phi_{i}\Phi_{j}+\frac{\lambda_{ijk}}{3}\Phi_{i}\Phi_{j}\Phi_{k})+ \text{h.c.}\\ &+ \int d^{4}\theta U_{ij}\bar\Phi_{i}\Phi_{j}+\int d^{2}\theta_{ij}(\chi\Phi_{i}\Phi_{j}+\eta_{ijk} \Phi_{i}\Phi_{j}\Phi_{k})+\text{h.c.}, 
\end{split}\end{equation}
where $U_{ij}=\mu_{s}^{2}u_{ij}\theta^{2}\bar{\theta}^{2}$, $\chi_{ij}=\mu_{s}^{2}x_{ij}\theta^{2}$ and $\eta_{ijk}=\mu_{s} n_{ijk}\theta^{2}$ are the soft breaking spurion superfields and $u,\ x,\ n$ are dimensionless matrices in the generations indices. We can write the resulting soft-breaking terms in components:
\begin{equation}\label{break1}
\mathcal{L}_{break}=\mu^{2}((u+x)_{ij}A_{i}A_{j}+(u-x)_{ij}B_{i}B_{j})+\mu n_{ijk}(A_{i}A_{j}A_{k}-3A_{i}B_{j}B_{k})\ .
\end{equation}
All the terms in (\ref{break1}) are super-renormalizable and they introduce new divergences in addition to the usual wave function renormalization of the Wess-Zumino model, which are studied in \cite{Girardello:1981wz, Iliopoulos:1974zv}. Since they all are super-renormalizable, these terms do not affect the divergent part of the wave function renormalization. Therefore, even for softly broken supersymmetry, it is easy to see that the Lorentz-violating parameter $c$ does not renormalize, because it does not appear explicitly in the superfield lagrangian. 
In the IR limit we need to fine tune $c$ according to the experimental bounds on Lorentz violation, but the low-energy value of $c$ will be also the value of the constant $c$ at high energy, because $c$ does not renormalize. Hence the deviation from the speed of light of the limiting speed of elementary particle is negligible for Lorentz-violating supersymmetric models. \newline
\indent 
We will consider an explicit model in components in order to understand how the behavior of the renormalization group equations is modified by the fact that the effective theory at low energies is the low-energy limit of a supersymmetric Lorentz-violating theory with soft breaking. 
For this specific model we will explicitly show  at one loop how the renormalization properties of a softly broken supersymmetric theory imply the non renormalization of $c$. Let us consider the most general low-energy limit of a renormalizable supersymmetric Lorentz-violating theory for an interacting chiral multiplet. Since supersymmetry is softly broken at low-energy, from the previous discussion it is clear that we can parametrize this breaking considering as independent the dimensionfull parameters of the lagrangian. Therefore the low-energy lagrangian will be described by six independent parameters: 
\begin{equation*}\begin{split}
\mathcal{L}&= \frac{(\hat\partial A)^2}{2}+\frac{c^{2}(\bar\partial A)^2}{2}+\frac{(\hat\partial B)^{2}}{2}+\frac{c^{2}(\bar\partial B)^2}{2}+\frac{m^{2}A^2}{2}+\frac{m^{\prime2}B^2}{2}+\frac{1}{2}\bar\psi(\hat{\slashed\partial}+v\bar{\slashed\partial}+M)\psi\\&+\frac{\lambda_{3}}{3!}A^{3}+\frac{\lambda_{3}}{2}^{\prime}AB^{2}+\frac{g^2}{4!}(A^{4}+B^{4}+6A^{2}B^{2})+gA\bar{\psi}\psi+igB\bar{\psi}\gamma_{5}\psi \ .
\end{split}\end{equation*}
In the approximation of small deviations from the speed of light we can put $c^{2}\simeq 1+\delta_{c^{2}}$ and $v\simeq 1+\delta_{v}$. The bare quantities are defined as:
\begin{align*}
&A_{b}=Z_{A}^{\frac{1}{2}}A\ ,\ \ \ \ B_{b}=Z_{B}^{\frac{1}{2}}B\ ,\ \ \ \ \psi_{b}=Z_{\psi}^{\frac{1}{2}}\psi \ ,\ \delta_{c^{2}b}= \delta_{c^{2}}+\Delta_{\delta_{c^{2}}}\ ,\  \delta_{c^{2}b}= \delta_{v}+\Delta_{\delta_{v}}\ , \\ 
&m^{2}_{b}=m^{2}+\Delta m^{2}\ ,\ \ \ \ m^{\prime2}_{b}=m^{\prime2}+\Delta m^{\prime2}_{B}\ ,\ \ \ \ M_{b}=M+\Delta M\ ,\\ 
&\lambda_{3b}=\mu^{\frac{\epsilon}{2}}(\lambda_{3}+\Delta_{\lambda_{3}})\ ,\ \lambda^{\prime}_{3b}=\mu^{\frac{\epsilon}{2}}(\lambda^{\prime}_{3}+\Delta_{\lambda^{\prime}_{3}})\ ,\ 
g_{b}=\mu^{\epsilon}(g+\Delta_{g})\ .
\end{align*}
Recalling the Feynman rules for Majorana fermions \cite{Denner:1992me} we obtain at one loop:
\begin{align*}
&Z_{A}=Z_{B}= 1-\frac{g^{2}}{2\pi^{2}\epsilon}(1-\bar{d}\delta_{v})\ ,\ Z_{\psi}=1-\frac{g^{2}}{2\pi^{2}\epsilon}(1-\frac{\bar{d}}{3}\delta_{c^{2}}-\frac{\bar{d}}{3}\delta_{v})\ ,\\
&\Delta_{\delta_{c^{2}}}=\frac{g^{2}}{2\pi^{2}\epsilon}(\frac{\delta_{c^{2}}}{2}-\delta_{v})\ ,\ \Delta_{\delta_{v}}= \frac{g^{2}}{2\pi^{2}\epsilon}(2\delta_{v}-\delta_{c^{2}})\ .
\end{align*}
If supersymmetry is softly broken the divergent part of the wave function renormalization has to be the same for all particles in the same supermultiplet. Therefore, imposing $Z_{\psi}=Z_{A}$ we obtain $2\delta_{v}=\delta_{c^{2}}$, which implies $\Delta_{\delta_{c^{2}}}=\Delta_{\delta_{v}}=0$. We thus conclude that the Lorentz-violating parameter does not renormalize. The only renormalization constant left is a common renormalization for the wave functions $Z_{A}=Z_{B}=Z_{\psi}= 1-\frac{g^{2}}{2\pi^{2}\epsilon}(1-\bar{d}\delta_{v})$, that at the zeroth order in $\delta_{v}$ is in agreement with the well known result for the Wess-Zumino model.\newline\indent 
An interesting problem to address concerns the nature of the constant which parametrizes the deviation from the speed of light in supersymmetric Lorentz-violating theories. In particular we want to understand whether or not the weighted constant $c$ is physically observable. As was already noted in \cite{GrootNibbelink:2004za}, if we consider a supersymmetric Lorentz-violating theory with one single sector the parameter $c$ appears both in the kinetic lagrangian (\ref{slv22}) and in the supersymmetry transformations (\ref{eq.slv0}) and can therefore be reabsorbed by a rescaling of the spatial coordinates $x^{\prime}_{\bar\mu}=cx_{\bar\mu}$. Now, as far as supersymmetry is an exact symmetry of our theory, all interacting supermultiplets have the same limiting speed $c$, because this parameter explicitly appears in the supersymmetry transformations. In this type of theories the parameter $c$  is physically unobservable because we can always set it to one by suitably choosing the length units. However, as was suggested in \cite{Colladay:2010tx}, we can construct more complicated situations with two or more sectors which are separately invariant under supersymmetry transformations (\ref{eq.slv0}) with different limiting speeds $c_{i}$, where the lower index labels the number of sectors. For example let us consider two different sectors $S_1$ and $S_2$, separately invariant under the supersymmetry transformations
\begin{align}
&\left\lbrace Q_{\alpha},\bar{Q}_{\dot{\alpha}}\right\rbrace= 2\sigma^{\hat{\mu}}_{\alpha\dot{\alpha}}P_{\hat{\mu}}+2c_{1}\sigma^{\bar{\mu}}_{\alpha\dot{\alpha}}P_{\bar{\mu}}\ ,\label{ss1}\\  
&\left\lbrace Q_{\alpha},\bar{Q}_{\dot{\alpha}}\right\rbrace= 2\sigma^{\hat{\mu}}_{\alpha\dot{\alpha}}P_{\hat{\mu}}+2c_{2}\sigma^{\bar{\mu}}_{\alpha\dot{\alpha}}P_{\bar{\mu}}\ .\label{ss2}
\end{align}
If $S_1$ and $S_2$ are completely decoupled, the supersymmetry algebras (\ref{ss1}, \ref{ss2}) are exactly realized in their respective sectors.\footnote{The Lorentz-violating theories that we are considering are rigid supersymmetric theories with very good approximation. Indeed the scale $\Lambda_{L}$ has to be around $10^{14}$ GeV in order to explain the neutrino masses \cite{Anselmi:2008bt}. Therefore we can neglect gravitational effects and consider the two sectors $S_{1}$ and $S_{2}$ completely decoupled.} We can still rescale the spatial coordinates in order to reabsorb $c_{1}$ or $c_{2}$ but, after the rescaling, we will have a Lorentz invariant sector $S1$ with $c_{1}\neq 1$ and an Lorentz violating hidden sector $S_2$ completely decoupled with $c_{2}\neq0$. Moreover, we can make $S_1$ and $S_2$ interacting by adding super-renormalizable interactions which will softly break supersymmetry in both sectors. The deviation from the speed of light in $S_2$ then becomes experimentally observable because any rescaling performed in order to remove the $c_{2}$ factor will produce Lorentz-violating effects in $S_1$, as was already pointed out in \cite{Colladay:2010tx}.\newline\indent
In conclusion, there is always the possibility to set one limiting speed to one by rescaling the spatial coordinates. This fact can make the Lorentz-violating supersymmetric models presented in \cite{Berger:2001rm, Colladay:2010tx} physically equivalent to the Lorentz invariant ones if we restrict our attention to models with a single sector. Besides that we have shown that the parameter $c$ does not renormalize in any softly broken supersymmetric theory, so that even if the deviation from the the speed of light is indeed observable it would be extremely small at any energy scale.

\section{Conclusion}
In this paper we have investigated the possibility of constructing supersymmetric Lorentz-violating theories that can be renormalized by a weighted power counting. Our analysis starts from the observation that supersymmetry and Lorentz violation are compatible at the level of the algebra~\cite{Berger:2001rm}.\newline\indent 
Moreover, we have shown that in the Lorentz-violating case it is possible to construct  new superalgebras with supercharges non-linear in the spatial momenta. However, the non-linearity of the supercharges makes the problem of finding interacting theories invariant under the new superalgebras very involved, because the superfield formalism looses its usefulness. As an example of this general difficulty we have shown that the interacting theory constructed in \cite{Xue:2010ih} is not invariant under this new class of superalgebras.\newline\indent 
Assuming linearity of the supercharges in the spatial momenta we find renormalizable supersymmetric Lorentz-violating models for even $n$ and classify them. It is straightforward to verify in the superspace formalism that the non-renormalization theorem is still valid in the Lorentz-violating case and as a consequence our models exhibit the improved ultraviolet behavior typical of supersymmetric theories. Moreover, the  weighted constant $c$ appearing in the supercharges algebra which parametrizes the limiting speed of the multiplet does not renormalize at high energies because the K\"ahler potential of a renormalizable Lorentz-violating theory has the same form as in the Lorentz invariant case. Furthermore, the low-energy recovery of Lorentz symmetry is parametrized only by this parameter $c$, which does not renormalize even at low energies if we assume supersymmetry to be broken softly.\newline\indent In the case of gauge theories we show that demanding supersymmetry implies that the weighted power counting has to coincide with the usual one. The only Lorentz-violating operators are then introduced by the weighted constant $c$, which does not renormalize and has to be very close to the speed of light at low energies in order to satisfy the experimental bounds on Lorentz violation \cite{Kostelecky:2008ts}. Therefore, if we demand renormalizabilty, gauge invariance and supersymmetry, the Lorentz invariance follows as a consequence.  \newline\indent Our analysis agrees with the conjecture that supersymmetry can solve the Lorentz fine tuning problem for Lorentz-violating theories, but at the same time it reveals that the requirement of supersymmetry restricts drastically the possibility of constructing renormalizable Lorentz-violating theories at high energies. Indeed, the final picture which emerges from our investigation is that the only possible models with non trivial Lorentz-violating operators involve neutral chiral superfields and do not have a gauge invariant extension. Therefore, if we want to construct renormalizable Lorentz-violating extensions of the Standard Model which have new interesting phenomenological consequences, the Lorentz fine tuning problem  \cite{Collins:2004bp, Jain:2005as} does not seem solvable by the requirement of supersymmetry.

\section*{Acknowledgements}
The author is grateful to D. Anselmi for suggesting the problem and for useful discussions. He also thanks R. Argurio for useful comments. The research of D.R. is supported by the ``Communaut\'e Fran\c{c}aise de Belgique'' through the ARC program, and also in part by IISN-Belgium (conventions 4.4511.06, 4.4505.86 and 4.4514.08) and by the Belgian Federal Science Policy Office through the Interuniversity Attraction Pole IAP VI/11.

\let\oldbibitem=\bibitem\renewcommand{\bibitem}{\filbreak\oldbibitem}
\bibliographystyle{utphys}
\fussy
\bibliography{BibliographySUSLV.bib}
\end{document}